\documentclass[sigconf, nonacm]{acmart}
\usepackage[english]{babel}
\usepackage{amsmath}
\usepackage{graphics}
\AtBeginDocument{%
	\providecommand\BibTeX{{%
			\normalfont B\kern-0.5em{\scshape i\kern-0.25em b}\kern-0.8em\TeX}}}
\begin{document}
	\title{Different Perspectives on FLP Impossibility}
	
	\author{Ivan Klianev}
	\email{Ivan.Klianev@gmail.com}
	\affiliation{%
	\institution{Transactum Pty Ltd}
	\city{Syndey}
	\country{Australia}
	}
	\begin{abstract}
		
	We demonstrate termination of binary consensus under the model and conditions used by Fischer, Lynch, and Patterson (FLP) to prove impossibility of binary agreement – in complete asynchrony and a possible process crash – in two steps. First, we introduce a new paradigm for consensus termination and show that impossibility of agreement is not evidence for impossibility to terminate. Next, we present a consensus algorithm that ensures termination with agreement about the initial input from the processes.
		
\end{abstract}

		

	\maketitle	
	
	\section{Introduction}
	
		The FLP impossibility result \cite{FLP_result} is about possible non-termination of every binary consensus protocol in asynchrony. 
		The theorem demonstrates the impossibility of binary agreement in complete asynchrony and a potential process crash. The proof of inability to terminate follows from the impossibility of agreement.
		
		We show that termination with binary agreement is an implicit requirement, rendered non-essential by the present state of art: i) linearisable (atomic) consistency \cite{Herlihy_Wing_1990} across multiple copies of the same data; and ii) state machine replication
		\cite{Lamport_StateMachine} 
		\cite{Lampson_1996}. 
		Further, we present a new paradigm for consensus termination, built on top of this art: i) the protocol terminates with agreement about the initial inputs; and ii) the processes individually compute the final agreement value. Thus, proving impossible termination of any consensus in asynchrony requires a proof of impossible termination with agreement about the initial input from the processes.
		
		We solved the impossibility to ensure termination of binary consensus in complete asynchrony and a possible crash. First, we demonstrated the correctness of the new paradigm by proving that consensus about a binary decision from a vector of initial values produces the same outcome as a binary decision from a consensus about a vector of initial values. Next, we designed an algorithm that solves vector consensus \cite{PeaseShostakLamport80} in asynchrony and a possible crash, and proved its unconditional termination. Thus, binary consensus terminates before the processes compute the agreement value.
			
	\subsection{Background}	
		
		Fischer, Lynch, and Patterson proved the FLP impossibility result in the following way: i) Presented a scenario where in asynchrony a consensus protocol cannot reach a decision state (i.e. terminate) with one crash. In this scenario, the correct processes wait infinitely long for a missing critical for decision state message, being unaware that what causes its missing is not asynchrony but inability of a crashed process to send it; ii) Proved that a protocol starting with randomly assigned 1-bit values and aiming to make a binary decision, with one unannounced crash-fail has at least one set of initial values that do not allow a binary decision, i.e., it is bivalent; and iii) Proved that a bivalent set of initial values may lead the protocol to a state where one missing message can prevent it from making binary decision, needed for reaching decision state. 
		The result states that if “consensus problem involves an asynchronous system of processes”, then “every protocol for this problem has the possibility of non-termination, even with only one faulty process”.
		
		\textbf{Conditionality of FLP Proof}
 
 		The proof of non-termination rests on the combined effect of two conditions: i) the contingency of agreement on the content of the initial values; and ii) a choice based on an implicit assumption.
		
		\textbf{Contingency of Agreement}
		
		Using the rules of arithmetic, it is impossible to always produce binary value from a random set of data. Counting the votes of an even number of voters for two candidates occasionally produces a tie. Regardless of how small the probability is for a tie, its possibility is certain. This arithmetical fact does not itself lead to consensus protocols’ possible non-termination, unless something within the protocols ensures that it does. This ‘something’ is a critical implicit assumption in the system model used to prove the FLP theorem. 
		
		\textbf{The Implicit Assumption}
		
		The assumption: a consensus protocol can terminate only with consensus on the agreement value. It transforms the \textit{inevitable} possibility for \textbf{tie} into an \textit{unavoidable} blocking of the protocol due to inability to choose the right action. If the tie is caused by a crash, waiting for the critical message will be infinite; if by a delayed message, a decision before receiving it may compromise safety. 
		A protocol that terminates with agreement on the initial inputs does not face this dilemma. Yet, when the impossibility result was shown, termination with a binary agreement had no known alternative.

		\subsection{Motivation}

		Binary consensus is critical for atomicity of third-party coordinated distributed transactions \cite{JimGray78} \cite{MohanStrongFink1985} \cite{HectorGM_EtAl1986}. 
		This coordination relies on a choice from a family of distributed commit protocols 
		\cite{JimGray78} 
		\cite{BruceLindsay79} 
		\cite{LampsonSturgis76}, each susceptible to blocking \cite{BernsteinHadzilacosGoodman}. Binary consensus non-termination is one of the reasons for blocking \cite{Gray_lamport}.
		
		In contrast, a decentralised database system operates with global replication of local transactions and ensures atomic consistency across local systems with state machine replication
		\cite{Lamport_StateMachine} 
		\cite{Schneider_1990} 
		\cite{Lampson_1996}.  
		Agreement for its purposes is about a vector containing state and the state value can in no way make the agreement impossible.

		Proving termination of binary consensus in complete asynchrony
		with tolerance to one crash would 
		naturally challenge the axiomatic acceptance of the influential, both in theory and practice, conjecture that no consensus protocol can operate with ensured simultaneous validity, safety, and liveness in partial synchrony \cite{GilbertLynch_2012}.
		
	\subsection{Problem Formulation}
	
		The solution requires solving two individual problems:
		
		1. Demonstrate that consensus about a binary decision from a vector $V=(v_1,...,v_n)$ produces the same outcome as a binary decision from a consensus about a vector $V=(v_1,...,v_n)$, i.e., $Bc(Bf(V)) = Bf(Vc(V))$, where $Bc$ denotes execution of deterministic \textit{binary} consensus protocol, $Vc$ denotes execution of deterministic \textit{vector} consensus protocol, $Bf$ is a deterministic function returning binary output
	
		2. Given that every process $P_i$, $i\in\{1,...,n\}$, has an initial value $v_i\in\{0,1\}$, demonstrate that no less than $ (n-1) $ processes can agree on an $n$-vector $ V $, where $V=(v_1,...,v_n)$, with at most one element replaced by a null marker $ \varnothing $, denoting an empty element.
	
	\subsection{Contribution}

		This work contributes to the subject of deterministic consensus:
		
		1. An entirely new paradigm for consensus termination.
			
		2. A consensus algorithm that tolerates a crash and terminates in complete asynchrony.

		3. Proof of possibility for consensus in asynchrony with a crash. 
		
		4. Evidence for possibility of simultaneous validity, safety, and liveness in complete asynchrony.
		
	\subsection{Content}		
 
		Section 2 shows the correctness of a new paradigm for consensus termination. Section 3 presents the system model. Section 4 introduces an algorithm with vector agreement in asynchrony and a crash. Section 5 proves the algorithm termination. Sections 6 and 7 illustrate the algorithm's phases. Section 8 shows termination of consensus with binary agreement. Section 9 contrasts our results with Crusader agreement. Section 10 concludes the paper.
	
	\section{New Paradigm for Termination}
	
		This section demonstrates that every consensus can terminate in two alternative ways: with agreement on consensus decision value or with agreement on the content of the initial values. The ability to terminate with agreement on the initial values shows that inability of agreement cannot be used as proof of inability to terminate.
		
		\textbf{Definition 1}: \textit{Vector Agreement.} A type of agreement where $ n $ processes $P_i$, $i\in\{1,...,n\}$ : a) start with individual initial value $v_i$ that can be a vector with finite number of elements; b) individually compose an $ n $-vector $V_i=(v_1,...,v_n)$; and c) terminate when no less than the required for quorum number of processes  $P_i$ agree on a vector $V=(v_1,...,v_n)$ containing at least the initial values $v_i$ of the agreed processes while the rest of vector elements can be replaced with a $ \varnothing $, denoting an empty element.
		
		Definition 1 is for vector agreement that was solved in synchrony \cite{PeaseShostakLamport80} and is good for the purposes of this paper. It excludes any agreement about a vector that is composed with interpretation of the content of the initial values or depends on their content. 
		
	\subsection{State Machine and Atomic Consistency}
	
		Replicated deterministic state machine \cite{Lamport_StateMachine} is the general way to implement a highly available system. It requires replicas to use
		a deterministic consensus algorithm \cite{PeaseShostakLamport80} to agree on each input. Thus, state machine replication is a way to arrange each replica, built as a deterministic state machine, to do exactly the same thing. Transition relation of a deterministic state machine is a function from (state, input) to (new state, output) \cite{Lampson_1996}. The replicas starting from exactly the same state and doing exactly the same thing finish with exactly the same state. While the state machine replication depends on a vector consensus algorithm as per Definition 1, the new paradigm for consensus termination relies on utilisation of dependence in the opposite direction. Under this paradigm, every consensus algorithm can terminate with a vector agreement on input from processes and every process can individually use state machine replication to compute the consensus decision value.

		Atomic consistency is also known as linearisable consistency \cite{Herlihy_Wing_1990} or atomic memory. For a system with data replicated on multiple computers, atomic consistency is what the sequential consistency \cite{Lamport_1979} is for a multiprocessor computer that concurrently executes multiple programs, each with update-access over the same data.
		Atomic consistency across multiple replicas is an equivalent to sequential consistency, yet with a critical constraint. On a single computer system, sequential consistency is correct with sequential execution under \textit{any} sequential order.
		Atomic consistency, however, requires every replica to perform the sequential execution under exactly \textit{the same} sequential order.
		
		After termination of a vector consensus protocol with ensured safety, the value of agreed vector is the same on each of the agreed computers, i.e., it is atomically consistent across these computers. Execution by the agreed computers of a function, involving the same sequence of computations over the agreed vector value, produces a consensus value that is atomically consistent across them.
	
	\subsection{Commutativity of Protocol Steps}
	
		The difference, between sequentially consistent data on a single computer and atomically consistent data across multiple replicas, explains the necessity for sequential ordering of the requested transactions for the purposes of state machine replication. With this in mind, the processes of a consensus system have to agree not just on each input, typically containing a multitude of requested transactions, but also establish a sequential order by placing them in a vector and agree on its value. The same ordering and agreement on a vector value is critical for the new paradigm for consensus termination. We claim that the most generic, and therefore the most fundamental consensus is the one with a vector agreement as per Definition 1. In support, we demonstrate that the agreement value of every consensus can be produced from a vector agreement.
	
		Let $S$ be a system of $N$ processes, $ N \geq 3 $, that operates with a deterministic consensus protocol. For the clarity of proof we assume synchronised synchronous processes, synchronous communication, and non-faulty links. Consensus requires $(N/2 +1)$ correct processes. 
		$S$ operates with a majority of $ C $ correct processes $(C \geq N/2 +1)$. The rest of the processes can be crash-failed.

		\textbf{Consensus Protocol Phases}

		\textbf{Phase 0}

		\textit{Step One}: Every process $P_{i}$ broadcasts a proposal $\Pi_{i}$ and every process $P_{j}$ receives $\Pi_{ij}$ from every $P_{i}$ or $\varnothing$, indicating a non-received proposal from a faulty process.

		\textit{Step Two}: Every process $P_{j}$ produces a vector $V_{j}$ that contains all received proposals $\Pi_{ij}$, including the  $\Pi_{jj}$ it received from itself, and  $\varnothing$ where a proposal was not received. 

		\textbf{Phase 1}

		\textit{Step One}:  Every process $P_{i}$ broadcasts a vector $V_{i}$ and every process $P_{j}$ receives a vector $V_{i}$ from every $P_{i}$ or $\varnothing$, indicating a non-received vector from a faulty process.

		\textit{Step Two}: System state is represented by a matrix of $ N $ rows and $ N $ columns. Row $R_{j}$ represents what process $P_{j}$ knows about the system state and contains $ N $ elements $E_{ji}$, where each element is either a vector $V_{ij}$ received from $P_{i}$ or from itself, or $\varnothing$. Process $P_{j}$ analyses the content of vectors in $R_{j}$, finds $C$ vectors $V_{ij}$ with $C$ elements $\Pi_{pq}$ having the same value, and constructs a decision vector $ V_{j} $ by including every $\Pi_{pq}$ with the same value in $C$ vectors $V_{ij}$ and assigns $\varnothing$ to the rest. As a result, $C$ processes $P_{j}$ construct a decision vector $ V_{j} $, where each $V_{j}$ contains the same vector value $ V $. Vector value $ V $ e critical for the proof.

		\textbf{Phase 2}  
		
		\textbf{Under the Traditional Paradigm for Termination:}

		\textit{Step One} (produces binary values): Every process $P_{i}$ executes a deterministic function \textit{Bf} passing $V_{i}$ as input and receives a binary value $B_{i}$.

		\textit{Step Two} (produces an agreed value): Every $P_{i}$ broadcasts $B_{i}$ and receives a $B_{j}$ from every $P_{j}$. In the matrix of system state, $C$ processes $P_{j}$ have an associate row $R_{j}$ with $C$ elements $E_{ji}$ containing the same $B_{i}$ value, which is the agreed binary value $ B $. 

		\textbf{Phase 2} 
		
		\textbf{Under the New Paradigm for Termination:}

		\textit{Step One} (produces an agreed value): Every process $P_{i}$ broadcasts vector $V_{i}$ and receives a vector $V_{j}$ from every $P_{j}$. In the matrix of the system state, $C$ processes $P_{j}$ have an associate row $R_{j}$ with $C$ elements $E_{ji}$ with the same $V_{i}$ value. This value is the agreed upon vector value $ V $.

		\textit{Step Two} (produces binary values): Every process $P_{i}$ executes the same deterministic function \textit{Bf} passing $V$ as input and receives a binary value $B$.

	\textbf{Proof of Commutativity}

		\textbf{Theorem 1}: \textit{The two steps of the final phase of a binary consensus protocol are commutative.}

		\textbf{Proof}

		Under the traditional paradigm, every correct process $P_{i}$ passes its vector $ V_{i} $ loaded at step One of Phase 1 with value $ V $ to \textit{Bf} and obtains a $ B_{i} $ loaded with the same binary value $ B $. Next, all correct processes agree on $ B $ as consensus outcome value. 

		Under the new paradigm, all correct processes $P_{i}$ agree on a vector value $ V $ as outcome. Next, every correct process $P_{i}$ passes value $ V $ to \textit{Bf} and obtains a binary value $ B $ as  consensus outcome.

		Under each paradigms, \textit{Bf} takes value $ V $ as input and produces the same value $ B $ as output. Also under each paradigms, inability to produce binary value $ B $ from vector value $ V $ prevents producing a binary consensus outcome. 	
		
	\subsection{Termination in Synchrony}

		Relevant to binary agreement in complete synchrony is the work of Santoro and Widmayer \cite{SantoroEtAl1989}. It  
		demonstrated that the dynamic transmission faults have the same negative effect on the possibility for binary agreement in asynchrony. This work made conclusions about the impossibility of binary agreement but wisely did not use the proven impossibility of agreement as basis to claim anything beyond that. Specifically, it does not claim impossibility of binary consensus based on the impossibility of agreement.
		
		Back to the commutativity of steps in synchrony, the fact that termination of vector consensus depends in no way on the content of the initial values hints that the impossibility of binary agreement cannot prevent termination. A protocol with design that implements the new paradigm separates termination from making the binary decision. It terminates and then obtains the consensus agreement value, if possible.	

		\textbf{Proof of Ensured Termination}

		\textbf{Corollary 1}: \textit{A binary consensus protocol can ensure termination in synchrony.}

		\textbf{Proof}

		The protocol can first terminate with an agreement on a vector value and then the processes can individually try to obtain the consensus outcome value with execution of a deterministic function [Theorem 1]. The proof of termination in synchrony follows from the proven termination \cite{PeaseShostakLamport80} of deterministic vector consensus in complete synchrony with all links being correct and process faults up to the tolerated fraction of all processes.

	\section{System Model}
		
		The previous section presented a new paradigm for termination of consensus protocols and proved it is as correct as the traditional paradigm. It also referred to an authoritative work in distributed computing where the inability of binary agreement in complete \textit{synchrony} is not used as proof of inability to terminate consensus. Similarly, stepping on the new paradigm for termination, we will argue that the inability to produce binary agreement in complete \textit{asynchrony} is not an evidence of inability to terminate consensus. 
		
		Proving possibility / impossibility to terminate consensus in complete asynchrony requires a proof of possibility / impossibility to terminate consensus with agreement about
		the values of initial input from processes of the consensus system, i.e., to terminate vector consensus as per Definition 1. The system model below is for the algorithm for vector consensus in complete asynchrony we are going to present. It is as close as possible to the model used to prove the FLP of impossibility of binary agreement in asynchrony.
		
		\textbf{The Model}
		
		A consensus protocol operates within a fully asynchronous system of $ N $ process $(N \geq 5)$. The protocol cannot make any assumptions about relative speed of processes, about delay time in delivering a message, or about order of delivery of messages, and also cannot detect the death of a process. So it is impossible for a process to tell whether another process has died (stopped entirely) or it is running very slowly.

		Each process $P$ has an \textit{input register} $R^{In}$, an \textit{output register} $R^{Out}$, and an unbounded amount of internal storage. Register $R^{Out}$ is a vector of $N$ elements $R^{In}$: $R^{Out}=( R^{In}_{1},R^{In}_{2},...,R^{In}_{N} )$.
		
		\textit{Internal state} of a process $ P $ comprises value in $R^{In}$ and value in $R^{Out}$, together with a program counter and internal storage.

		\textit{Starting values} of the internal state are fixed, with an exception for the starting value in $R^{In}$. More specifically, $ R^{In} $ starts with a \textit{length prefixed string} having an arbitrary value from a finite universe $ S $, and $ R^{Out} $ starts always with the fixed value:

		$R^{Out}=( \varnothing,\varnothing,...,\varnothing )$, 
		where $ \varnothing $ is a null marker denoting an empty element.
		
		Objective of the consensus protocol is to establish an agreement between $\geq (N-1)$ correct processes in the possibility of one crash.

		\textit{Proposal} is a vector $\Pi$ of $ N $ elements that contains at most 1 element with value $\varnothing $.

		\textit{Decision} is a vector of $ N $ elements that contains at most 1 element with value $\varnothing $. Every process broadcasts a decision after it has \textbf{received} $ (N-2) $ equal agreement proposals.

		\textit{Decision state} is an internal state a process $ P $ reaches
		after it has broadcast a decision and has collected $ (N-1) $ equal decisions. Once a decision state is reached, $P$ writes its value in register $ R^{Out} $.

		\textit{Decision value} is the actual vector value $V^{De}$ which $P$ writes in $R^{Out}$ after reaching a decision state.

		$ P $ acts deterministically according to a transition function, which cannot change the value of $R^{Out}$ once $ P $ reaches a decision state. The system is specified by the transition function $ \Phi $ associated with each of the processes and the initial values in the input registers.	The processes strictly follow function $ \Phi $, i.e., no Byzantine behaviour.

		The system operates with no faulty links.
		Processes communicate by sending each other messages. 
		A \textit{message} is a tuple $(P_{s}, P_{d}, m)$, where $P_{s}$ is the sender process,  $ P_{d} $ is the destination process, and $ m $ is a 'message value' from a fixed universe $ M $. The message system maintains a \textit{message buffer}.
		Message buffer is a multi-set of messages that have been sent but not yet delivered. It supports two abstract operations:

		- \textit{Send}($P_{s}, P_{d}, m$). It places ($P_{s}, P_{d}, m$) in the message buffer.

		- \textit{Receive}($P_{d}$). It deletes some message ($P_{s}, P_{d}, m$) from the buffer and returns $ m $, in which case ($P_{s}, P_{d}, m$) is \textit{delivered}, or $ \varnothing $ and leaves the buffer unchanged.

		Thus, the message system acts non-deterministically, subject only to the condition that if \textit{Receive}($P_{d}$) is performed infinitely many times, then every message ($P_{s}, P_{d}, m$) in the message buffer is delivered. In particular, the message system is allowed to return $ \varnothing $ a finite number of times in response to \textit{Receive}($P_{d}$), even though a message ($P_{s}, P_{d}, m$) is present in the buffer.

		Basic terms:

		- \textit{Configuration}. A set containing the internal state of every system process and the contents of its message buffer.

		- \textit{Step}. A step takes one configuration to another and consists of a primitive step by a single process $ P $. A step executes a sequence of activities as one atomic operation. Let $ C $ be a configuration. A step occurs in two phases. First, \textit{Receive}($P$) on the message buffer in $ C $ obtains a value $m \in M \cup |\varnothing|$. Next, depending on the internal state of $ P $ in $ C $ and on value $ m $, process $ P $ may enter a new internal state and may send a finite set of messages to other processes. Since the processes are deterministic, a step of $ P $ is completely determined by the internal state of $ P $ and the received value $ m $.

		- \textit{Event}. A tuple ($P_{s}, P_{d}, m$) indicating delivered $ m $ from $P_{s}$ to $P_{d}$.

		- \textit{Schedule}. A schedule from configuration $ C $ is a finite or infinite sequence of events that can be applied in turns, starting from $ C $. For a process $ P $ schedules are commutative \cite{FLP_result}, until $ P $ sends a $\Pi$.

		- \textit{Run}. A sequence of steps taken in association with a schedule.

		- \textit{Correct process}. A process is correct in a run if it takes infinitely many steps. Otherwise it is \textit{faulty}.

		Configurations:

		- \textit{Initial}. A configuration in which each process starts at the initial state and the message buffer is empty.

		- \textit{Resulting}. A configuration that is a result from execution of any schedule, regardless of the number of events.

		- \textit{Reachable}. A configuration is reachable if it is a result from execution of a schedule with a finite number of events.

		- \textit{Accessible}. A configuration is accessible if it is reachable from some initial configuration.

		- \textit{Decision}. An accessible configuration where at least $ (N-1) $ processes have reached decision state.

		Run types:

		- \textit{Admissible run}. A run starting from an initial configuration where at most one process is faulty and all messages sent to correct processes are received.

		- \textit{Deciding run}. An admissible run that has reached a decision configuration. 

		Correctness Condition:
		
		- \textit{Correct protocol}. A consensus protocol is correct if every admissible run is a deciding run with one decision state.

	\section{Vector Consensus in Asynchrony}	
	
		In this section we present an algorithm that implements consensus with vector agreement under the defined model. The algorithm ensures that every admissible run is a deciding run (liveness) and all correct processes reach the same decision state (safety).

	\subsection{Algorithm Overview}

		The algorithm operates in 3 phases: i) Initial; ii) Proposals; and iii) Decision. From start to termination it broadcasts 4 messages: i) initial value; ii) first proposal; iii) second proposal; and iv) decision seed.
		
		\textbf{Initial Phase} 

		During this phase, the processes distribute their \textit{initial value} messages. A process $P_i$ starts this phase by broadcasting an \textit{initial value} message, i.e., $v_i$ loaded with the content of $R_i^{In}$. In a system of N processes, this phase completes after $P_i$ has received an \textit{initial value} message from $ (N-2) $ processes. An \textit{initial value} message received after completion of this phase is recorded in memory but ignored.

		\textbf{Proposals Phase} 

		During this phase, the processes distribute and process their \textit{first proposal}, i.e. $\Pi^{Frst}$, and \textit{second proposal}, i.e. $\Pi^{Scnd}$, messages. A process $P_i$ starts this phase by broadcasting $\Pi_i^{Frst}$ and completes it with a vector $V_i$ = $V^F$, i.e. having no $\varnothing$ element, or with $V_i$ = $V^{\varnothing}$ that can have $N$ different versions, each with a $\varnothing$ element on a different position. This phase narrows  the set of possible $V_i$ values in the system from $(N+1)$ to just 2, i.e., one version of $V^{\varnothing}$ and $V^F$.

		\textbf{Decision Phase} 

		During this phase, the processes distribute their \textit{decision seed} messages, denoted as $\Delta$ message. $P_i$ starts this phase by  broadcasting a $\Delta_i$ message containing $V_i$. Some processes start this phase with the same \ $V^{\varnothing}$, others start it with $V^F$. Regardless of that, all processes decide the same $V^{De}$ from the collected $\Delta$ messages. 	
		
	\subsection{Order of Message Processing}

		Under communication asynchrony the processes do not start a consensus round simultaneously and messages are not necessarily received by the order they are sent. A process records in memory the messages related to a phase that has not started yet and once the phase starts it begins the processing, normally in the order of receiving. 
		Within a phase, the sequential order of receiving is the exact sequential order of processing in regard only to \textit{first proposal} and \textit{decision seed} messages.
		The relation between the order of receiving and the order of processing of \textit{second proposal}, however, is regulated by the Order Rule.
		
		\textbf{Order Rule}: The order of processing of the messages from a sender is according to the order of sending.
		
		The order of sending is known to the receivers.  
		So, if the \textit{second proposal} message of process $P_i$ is received by process $P_j$ before $P_j$ has received the \textit{first proposal} message of $P_i$, the Order Rule requires process $P_j$ to record the received \textit{second proposal} message in memory and start processing it after the \textit{first proposal} message of $P_i$ has been received and processed. Thus the Order Rule reverses one of the effects of communication asynchrony.

	\subsection{Proposals Phase Rules}
		
		A process $P_i$ starts this phase by preparing and broadcasting its \textit{first proposal} message, i.e. $\Pi_i^{Frst}$. Later, $P_i$ may prepare and broadcast a \textit{second proposal} message, i.e. $\Pi_i^{Scnd}$. The difference between the two messages is that a $\Pi^{Frst}$ \textbf{always contains} a $ \varnothing $ vector element while a $\Pi^{Scnd}$ \textbf{never contains} a $ \varnothing $ vector element. 
		
		If $P_i$ has not yet broadcast a $\Pi_i^{Scnd}$ message, two rules regulate the preparing and broadcast of a $\Pi_i^{Scnd}$:
		
	 	- \textbf{Blend Rule 1}:
	 	Process $P_i$, having received from a process $P_j$ a $\Pi_j^{Frst}$ that contains a $ \varnothing $ vector element on a position with different index compared to the position index of the $ \varnothing $ vector element in $\Pi_i^{Frst}$, prepares and broadcasts a $\Pi_i^{Scnd}$ message, even after $P_i$ has completed Proposals Phase.
	 	
	 	- \textbf{Blend Rule 2}:
	 	Process $P_i$, having received and processed a $\Pi_j^{Scnd}$ message from $P_j$, prepares and broadcasts a $\Pi_i^{Scnd}$ message, even after $P_i$ has completed Proposals Phase.
	 	
	 	Completion of this phase is also regulated by two rules:
	 	
	 	- \textbf{Completion Rule 1}: Process $P_i$ completes this phase with a vector value $V_i$ = $V^\varnothing$ after it has received $(N-2)$ equal $\Pi^{Frst}$ messages. 
	 	
	 	- \textbf{Completion Rule 2}: Process $P_i$ completes with a vector value $V_i$ = $V^F$ after it has received and processed $(N-2)$ messages $\Pi^{Scnd}$.
	 	
	 	Proposals Phase ensures 3 critical certainties: i) Every correct process completes it with either $V^\varnothing$ or $V^F$; ii) All processes, completing with $V^\varnothing$, complete it with the same version of $V^\varnothing$; and iii) Configuration where only one process completes it with $V^F$ is not feasible. The part of the algorithm that implements the operation of Proposals Phase is illustrated in detail in Section 6.

	\subsection{Decision Phase Rule}
	 
	 	A process $P_i$ starts this phase by broadcasting a $\Delta_i$ message that contains $V_i$. Having received $(N-2)$ $\Delta$ messages, if $V_i$ = $V^{\varnothing}$ and one of these messages contains a $V^F$ vector, $P_i$ applies the Update Rule.
	 
	 	\textbf{Update Rule}: Starting Decision Phase with $V^{\varnothing}$, a process updates it to $V^F$ on receiving a $\Delta$ message containing $V^F$. 
	 
	 	Process $P_i$ completes Decision Phase after it has received $(N-2)$ $\Delta$ messages and has applied the Update Rule if the conditions made it necessary. $P_i$ completes with a vector value $V^F$ if it started the phase with $V^F$ or if the Update Rule was applied. Otherwise, $P_i$ completes the phase with a vector value $V^{\varnothing}$. The algorithm instance running on process $P_i$ terminates on completion of Decision Phase. The part of the consensus algorithm that implements the operation of Decision Phase is illustrated in detail in Section 7.

	\section{Proofs}

		\textbf{Theorem 2}: \textit{A vector consensus protocol can be correct in complete asynchrony in spite of a possible process crash}.

		\textbf{Proof}

		Consensus requires every process to distribute to every other process the data in its $R^{In}$ and all correct processes agree about ordering of individual processes' data in a single dataset. Sufficient is to show the ability to terminate reaching a decision configuration with only one decision state. The proof follows from these five lemmas:

		\textbf{Lemma 1}: \textit{No initial configuration can prevent correct processes from reaching a decision state.}

		\textbf{Proof}

		A \textit{decision run} of a process $ P $ starts from its initial state, involves receiving messages containing the initial state from at least $(N-2)$ other processes, and reaches a decision state after execution of a finite sequence of steps by $ P $ and by at least $(N-2)$ other processes. As the agreement is about an ordered set with the initial state data, not about a decision based on that data, reaching a decision state by $ P $ in no way depends on the content of the initial state data. 

		\textbf{Lemma 2}: \textit{A protocol can ensure that every non-crashed process enters Decision Phase.}

		\textbf{Proof}	

		Entering Decision Phase requires completion of Proposals Phase. 	If $\ge 2 $ processes obeyed Order Rule and completed Proposals Phase under Completion Rule 1 after each has received $(N-2)$ equal $\Pi^{Frst}$ messages, then the rest of processes will complete Proposals Phase under Completion Rule 2. Reasoning: The definitions of Blend Rule 1 and Blend Rule 2 explicitly obligate every process to broadcast a $\Pi^{Scnd}$ message, even after completion of Proposals Phase, if the conditions for one of the Blend Rules are presented and a $\Pi^{Scnd}$ message has not been broadcast. Thus, every correct process broadcasts a $\Pi^{Scnd}$ under Blend Rule 1 or Blend Rule 2.

		\textbf{Lemma 3}: \textit{A protocol can ensure that every correct process starts deciding either with all initial inputs or lacking the same initial input.}

		\textbf{Proof}

		A process completes Proposals Phase with a $V^\varnothing$ vector when it has received and processes ($N-2)$ equal $\Pi^{Frst}$ messages, i.e., each containing a $V^\varnothing$ vector having a $\varnothing$ element with the same index. In a system of $N$ processes, where $N \ge 5$, this itself precludes the possibility for existence of another ($N-2)$ processes that have broadcast equal $\Pi^{Frst}$ messages with a different $V^\varnothing$ vector.

		\textbf{Lemma 4}: \textit{A protocol can ensure that either more than one process or none starts deciding with a vector having all initial inputs.}

		\textbf{Proof}

		Assume existence of configuration where only one process $P_i$ enters Decision Phase with vector $V^F$ and the rest of processes enter Decision Phase with vector $V^\varnothing$.  

		\textit{Case with No Process Crash.} (see Section 6, Case 2 for details)
		
		The assumption means that $(N-1)$ processes completed Proposals Phase with vector $V^\varnothing$ under Completion Rule 1, which requires $(N-1)$ processes to broadcast equal $\Pi^{Frst}$ messages. Under Order Rule, process $P_i$ will respond to $(N-2)$ equal messages $\Pi^{Frst}$ before responding to $(N-2)$ messages $\Pi^{Scnd}$ and will also complete with vector $V^\varnothing$ under Completion Rule 1. Hence the assumption is incorrect.  
		
		\textit{Case with One Process Crash} (see Section 6, Case 4, Case 5 and Case 7 for details)
		
		The assumption means that $(N-2)$ processes completed Proposals Phase with vector $V^\varnothing$ under Completion Rule 1, which requires $(N-1)$ processes to broadcast equal $\Pi^{Frst}$ messages. When this requirement is satisfied (Subcase 1), these $(N-2)$ processes will complete with vector $V^\varnothing$ under Completion Rule 1; otherwise (Subcase 2) with vector $V^F$ under Completion Rule 2. In both, Subcase 1 and Subcase 2, process $P_i$ completes with vector $V^\varnothing$ under Completion Rule 1. Hence the assumption is incorrect. 

		\textbf{Lemma 5}: \textit{A protocol can terminate with all correct processes agreed either on all initial inputs or on lacking the same initial input.}

		\textbf{Proof} 	

		Decision Phase may have the following configurations:

		- \textit{Case A}: Only one process started with $V^F$. This possibility is precluded [Lemma 4].
		
		- \textit{Case B}: $\ge 2 $ processes started with different  $V^{\varnothing}$ vectors. This possibility is precluded [Lemma 3].

		- \textit{Case C}: $\ge 2 $ processes started with $V^F$. All correct processes broadcast a $\Delta$ message. Each one that started with $V^{\varnothing}$ receives one or more message containing a $V^F$ vector and, obeying the Update Rule, updates its vector from $V^{\varnothing}$ to $V^F$. Thus every correct process, knowing that every other correct process either started with $V^F$ or has updated to $V^F$, terminates with $V^{De}$ = $V^F$.

		- \textit{Case D}: All processes started with $V^{\varnothing}$. All correct processes broadcast a $\Delta$ message and no correct process receives a message with vector $V^F$. Thus, knowing that Decision Phase cannot start with less than two processes with vector $V^F$, every correct process knows that all correct processes have started with a $V^{\varnothing}$ and terminates with $V^{De}$ = $V^{\varnothing}$.		
		

	\section{The Algorithm – Proposals Phase}

		In a system $S$ of $ N $ processes, $ N \geq 5 $, let $P_{s}$ be a process, which is either slow or with slow outbound links, and let $P_{r}$ be the one that	affects the decision state depending on the sequence it receives the \textit{initial values} of the rest of processes.

		Let $(N-2)$ processes broadcast $\Pi^{Frst}$ with the same vector $V^{\varnothing}$, which includes the \textit{initial value} of $P_{r}$ and $\varnothing$ instead of the \textit{initial value} of $P_{s}$. Let also $P_{r}$ received $(N-3)$ \textit{initial values}. Receiving one more \textit{initial value} will allow $P_{r}$ to prepare and broadcast $\Pi_{r}^{Frst}$. 

	\subsection{Cases with No Crash}

		\textbf{Case 1:}
		
		At start: i) $ (N-2) $ processes have broadcast $\Pi^{Frst}$ with a $ \varnothing $ vector element instead of the \textit{initial value} $v_s$ of $P_{s}$; 
		ii) $P_{r}$ has received $v_s$ \textbf{before} broadcasting $\Pi_r^{Frst}$. Hence $\Pi_r^{Frst}$ contains $\varnothing $ vector element on position different from the position of $ \varnothing $ vector element in the \textit{first proposal} messages broadcast at (i);
		iii) $P_s$ broadcast $\Pi_s^{Frst}$ after receiving any $(N-2)$ \textit{initial values}. 
		
		All $N$ processes, obeying Blend Rule 1 or Blend Rule 2 broadcast a $\Pi^{Scnd}$. $P_{r}$ and $P_s$, obeying Order Rule and Completion Rule 1, complete with $V^\varnothing$. The rest of the processes complete with $V^F$.

		\textbf{Illustration}: Configuration with $P_5$ as $P_s$ and $P_4$ as $P_r$.

		$P_1$ : $\Pi_1^{Frst}$ = $( v_1, v_2, v_3, v_4, \varnothing )$.

		$P_2$ : $\Pi_2^{Frst}$ = $( v_1, v_2, v_3, v_4, \varnothing )$.

		$P_3$ : $\Pi_3^{Frst}$ = $( v_1, v_2, v_3, v_4, \varnothing )$.

		$P_4$ : $\Pi_4^{Frst}$ = $( v_1, v_2, \varnothing, v_4, v_5 )$.

		$P_5$ : $\Pi_5^{Frst}$ = $( v_1, v_2, v_3, \varnothing, v_5 )$.

		Obeying Blend Rule 1, 
		$P_1$ broadcasts $\Pi_1^{Scnd}$ on receiving $\Pi_4^{Frst}$ or $\Pi_5^{Frst}$, 
		$P_2$ broadcasts $\Pi_2^{Scnd}$ on receiving $\Pi_4^{Frst}$or $\Pi_5^{Frst}$, 
		$P_3$ broadcasts $\Pi_3^{Scnd}$ on receiving $\Pi_4^{Frst}$or $\Pi_5^{Frst}$, 
		$P_4$ broadcasts $\Pi_4^{Scnd}$ on receiving $\Pi_1^{Frst}$ or $\Pi_2^{Frst}$ or $\Pi_3^{Frst}$ or $\Pi_5^{Frst}$, and 
		$P_5$ broadcasts $\Pi_5^{Scnd}$ on receiving $\Pi_1^{Frst}$ or $\Pi_2^{Frst}$ or $\Pi_3^{Frst}$ or $\Pi_4^{Frst}$.
		
		Obeying Order Rule, 
		$P_4$ and $P_5$ collect $(N-2)$ equal $\Pi^{Frst}$ before $(N-2)$ $\Pi^{Scnd}$.
		
		Obeying Completion Rule 1, 
		$P_4$ and $P_5$ complete with $V^\varnothing$.
		
		Obeying Completion Rule 2, 
		$P_1$, $P_2$, and $P_5$ complete with $V^F$.

		\textbf{Case 2:}
		
		At start: i) $ (N-2) $ processes have broadcast $\Pi^{Frst}$ with a $ \varnothing $ vector element instead of the \textit{initial value} $v_s$ of $P_{s}$; 
		ii) $P_{r}$ has received $v_s$ \textbf{after} broadcasting $\Pi_r^{Frst}$. Hence $\Pi_r^{Frst}$ contains a $\varnothing $ vector element in the same position as the $ \varnothing $ vector element in the \textit{first proposals} broadcast at (i).
		iii) $P_s$ broadcast $\Pi_s^{Frst}$ after receiving any $(N-2)$ \textit{initial values}. 		
		
		All $N$ processes, obeying Order Rule and Blend Rule 1, broadcast $\Pi^{Scnd}$ messages. All $N$ processes, obeying Order Rule, collect $N-2$ equal $\Pi^{Frst}$ messages and complete with  $V^\varnothing$.
		
		\textbf{Illustration}: Configuration with $P_5$ as $P_s$ and $P_4$ as $P_r$.

		$P_1$ : $\Pi_1^{Frst}$ = $( v_1, v_2, v_3, v_4, \varnothing )$.

		$P_2$ : $\Pi_2^{Frst}$ = $( v_1, v_2, v_3, v_4, \varnothing )$.

		$P_3$ : $\Pi_3^{Frst}$ = $( v_1, v_2, v_3, v_4, \varnothing )$.

		$P_4$ : $\Pi_4^{Frst}$ = $( v_1, v_2, v_3, v_4, \varnothing )$.

		$P_5$ : $\Pi_5^{Frst}$ = $( v_1, v_2, v_3, \varnothing, v_5 )$.

		Obeying Order Rule and Blend Rule 1: 
		$P_1$, $P_2$, $P_3$, and $P_4$ broadcast $\Pi^{Scnd}$ on receiving  $\Pi_5^{Frst}$; and  
		$P_5$ broadcasts $\Pi^{Scnd}$ on receiving $\Pi^{Frst}$ from $P_1$, $P_2$, $P_3$, or $P_4$.	
		
		Obeying Order Rule, 
		$P_1$, $P_2$, $P_3$, $P_4$, and $P_5$ individually have $(N-2)$ equal $\Pi^{Frst}$ before $(N-2)$ $\Pi^{Scnd}$.
		
		Obeying Completion Rule 1, 
		$P_1$, $P_2$, $P_3$, $P_4$, and $P_5$ individually complete with $V^\varnothing$.

		\subsection{Cases with A Crash}
		
		Let $P_{c}$ be the crashed process in the system S. The effect of the crash is defined mainly by which stage of broadcasting $P_{c}$ was at the point of time when the crash happened.

		\textbf{Case 3}: Process $P_{c}$ crashed \textbf{before} broadcasting \textbf{\textit{initial value}}. The rest of processes wait for $R^{In}$ from $ (N-2) $ processes, then use the same set of data to build and broadcast the same $\Pi^{First}$ messages. All correct processes complete this phase with value  $V^{\varnothing}$. 
				
		\textbf{Case 4}: Process $P_{c}$ crashed \textbf{while} broadcasting \textbf{\textit{initial value}}. 

		\textbf{Scenario 4.1}: One process $P_{r1}$ received the \textbf{\textit{initial value}} of $P_{c}$ before entering Proposals Phase.	
		
		\textbf{Illustration}: Configuration with $P_5$ as $P_c$ and $P_4$ as $P_{r1}$.

$P_1$ : $\Pi_1^{Frst}$ = $( v_1, v_2, v_3, v_4, \varnothing )$.

$P_2$ : $\Pi_2^{Frst}$ = $( v_1, v_2, v_3, v_4, \varnothing )$.

$P_3$ : $\Pi_3^{Frst}$ = $( v_1, v_2, v_3, v_4, \varnothing )$.

$P_4$ : $\Pi_4^{Frst}$ = $( v_1, v_2, \varnothing, v_4, v_5 )$.

$P_5$ : $\Pi_5^{Frst}$ = $( v_1, v_2, v_3, \varnothing, v_5 )$.
		
		Obeying Order Rule, 
		$P_4$ have $(N-2)$ equal $\Pi^{Frst}$ before $(N-2)$ $\Pi^{Scnd}$ and, obeying Completion Rule 1, completes with $V^\varnothing$.
		
		Obeying Blend Rule 1,
		$P_1$, $P_2$, $P_3$, and $P_4$ individually broadcast $\Pi^{Scnd}$.
		
		Obeying Completion Rule 2, 
		$P_1$, $P_2$, and $P_3$ individually complete with $V^F$.
		
		\textbf{Scenario 4.2}: Less than $(N-2)$ processes received the \textbf{\textit{initial value}} of $P_{c}$ before entering Proposals Phase.
		
		\textbf{Illustration}: Configuration with $P_5$ as $P_c$, $P_4$ as $P_{r1}$, and $P_3$ as $P_{r2}$.

$P_1$ : $\Pi_1^{Frst}$ = $( v_1, v_2, v_3, v_4, \varnothing )$.

$P_2$ : $\Pi_2^{Frst}$ = $( v_1, v_2, v_3, v_4, \varnothing )$.

$P_3$ : $\Pi_3^{Frst}$ = $( v_1, v_2, \varnothing, v_4, v_5 )$.

$P_4$ : $\Pi_4^{Frst}$ = $( v_1, v_2, \varnothing, v_4, v_5 )$.

$P_5$ : $\Pi_5^{Frst}$ = $( v_1, v_2, v_3, \varnothing, v_5 )$.

		Obeying Order Rule and Blend Rule 1, 
		$P_1$, $P_2$, $P_3$, and $P_4$ individually broadcast $\Pi^{Scnd}$.
		
		Obeying Completion Rule 2, 
		$P_1$, $P_2$, $P_3$, and $P_4$ individually complete with $V^F$.
		
		\textbf{Scenario 4.3}:  $\ge (N-2)$ processes received the \textbf{\textit{initial value}} of $P_{c}$ before entering Proposals Phase.		
		
		\textbf{Illustration}: Configuration with $P_5$ as $P_c$, $P_4$ as $P_{r1}$, $P_3$ as $P_{r2}$, and $P_2$ as $P_{r3}$.

$P_1$ : $\Pi_1^{Frst}$ = $( v_1, v_2, v_3, v_4, \varnothing )$.

$P_2$ : $\Pi_2^{Frst}$ = $( v_1, \varnothing, v_3, v_4, v_5 )$.

$P_3$ : $\Pi_3^{Frst}$ = $( v_1, v_2, \varnothing, v_4, v_5 )$.

$P_4$ : $\Pi_4^{Frst}$ = $( v_1, v_2, \varnothing, v_4, v_5 )$.

$P_5$ : $\Pi_5^{Frst}$ = $( v_1, v_2, v_3, \varnothing, v_5 )$.

		Obeying Order Rule and Blend Rule 1, 
		$P_1$, $P_2$, $P_3$, and $P_4$ individually broadcast $\Pi^{Scnd}$.

		Obeying Completion Rule 2, 
		$P_1$, $P_2$, $P_3$, and $P_4$ individually complete with $V^F$.

		\textbf{Case 5}: Process $P_{c}$ crashed \textbf{after} broadcasting \textbf{\textit{initial value}}. 
		If all processes that received the \textit{initial value} of $P_{c}$ were already in Proposals Phase, All $(N-1)$ correct processes complete this phase with value  $V^{\varnothing}$.
		If at least one process $P_{r}$ has received the \textit{initial value} of $P_{c}$ before entering Proposals Phase, obeying Blend Rule 1 and Blend Rule 2, all correct processes complete with value $V^F$.

		\textbf{Case 6}: Process $P_{c}$ crashed \textbf{before} broadcasting $\Pi_c^{Frst}$. All $(N-1)$ correct processes complete this phase with value  $V^{\varnothing}$. 
		
		\textbf{Case 7}: Process $P_{c}$ crashed \textbf{while} broadcasting $\Pi_c^{Frst}$.
		
		\textbf{Scenario 7.1}: One process $P_r$ received the \textit{initial value} of $P_{c}$ before entering Proposals Phase.
		
		At start: 
		i) Process $P_r$ broadcast $\Pi_r^{Frst}$ containing the \textit{initial value} of $P_{c}$ and a $ \varnothing $ vector element on other position; and
		ii) $ (N-2) $ processes broadcast $\Pi^{Frst}$ with a $ \varnothing $ vector element instead of the \textit{initial value} of $P_{c}$.
		
		\textbf{Illustration}: Configuration with $P_1$ as $P_c$ and $P_2$ as $P_r$.
	
		$P_1$ : $\Pi_1^{Frst}$ = $( v_1, v_2, v_3, v_4, \varnothing )$.
	
		$P_2$ : $\Pi_2^{Frst}$ = $( v_1, v_2, \varnothing, v_4, v_5 )$.
	
		$P_3$ : $\Pi_3^{Frst}$ = $( \varnothing, v_2, v_3, v_4,  v_5 )$.
	
		$P_4$ : $\Pi_4^{Frst}$ = $( \varnothing, v_2, v_3, v_4, v_5 )$.
	
		$P_5$ : $\Pi_5^{Frst}$ = $( \varnothing, v_2, v_3, v_4, v_5 )$.
		
		Obeying Order Rule and Completion Rule 1, 
		$P_2$ completes with $V^\varnothing$.
		
		Obeying Blend Rule 1,
		$P_2$ broadcasts $\Pi_2^{Scnd}$ on receiving $\Pi_3^{Frst}$ or $\Pi_4^{Frst}$ or $\Pi_5^{Frst}$, 
		$P_3$ broadcasts $\Pi_3^{Scnd}$ on receiving $\Pi_2^{Frst}$ or $\Pi_5^{Frst}$,
		$P_4$ broadcasts $\Pi_4^{Scnd}$ on receiving $\Pi_2^{Frst}$ or $\Pi_5^{Frst}$, and
		$P_5$ broadcasts $\Pi_5^{Scnd}$ on receiving $\Pi_2^{Frst}$ or $\Pi_3^{Frst}$ or $\Pi_4^{Frst}$.
		
		Obeying Completion Rule 2:
		$P_2$ completes with $V^F$ on receiving $\Pi_3^{Scnd}$, $\Pi_4^{Scnd}$, and $\Pi_5^{Scnd}$;
		$P_3$ completes with $V^F$ on receiving $\Pi_2^{Scnd}$, $\Pi_4^{Scnd}$, and $\Pi_5^{Scnd}$;
		$P_4$ completes with $V^F$ on receiving $\Pi_2^{Scnd}$, $\Pi_3^{Scnd}$, and $\Pi_5^{Scnd}$; and
		$P_5$ completes with $V^F$ on receiving $\Pi_2^{Scnd}$, $\Pi_3^{Scnd}$, and $\Pi_4^{Scnd}$.
		
		\textbf{Scenario 7.2}: $ > $ 1  but $ < (N-2) $ processes received the \textit{initial value} of $P_{c}$ before they entered Proposals Phase.

		At start: 
		i) Process $P_{r1}$ broadcast $\Pi_{r1}^{Frst}$ containing the \textit{initial value} of $P_{c}$ and a $ \varnothing $ vector element on other position; 
		ii) $ (N-3) $ processes broadcast $\Pi^{Frst}$ with a $ \varnothing $ vector element instead of the \textit{initial value} of $P_{c}$; and 
		iii) Process $P_{r2}$ broadcast $\Pi_{r2}^{Frst}$ with a $\varnothing $ vector element on a position different from the positions in (i) and (ii). 

		\textbf{Illustration}: Configuration with $P_1$ as $P_c$, $P_2$ as $P_{r1}$, and $P_5$ as $P_{r2}$.

$P_1$ : $\Pi_1^{Frst}$ = $( v_1, v_2, v_3, v_4, \varnothing )$.

$P_2$ : $\Pi_2^{Frst}$ = $( v_1, v_2, \varnothing, v_4, v_5 )$.

$P_3$ : $\Pi_3^{Frst}$ = $( \varnothing, v_2, v_3, v_4,  v_5 )$.

$P_4$ : $\Pi_4^{Frst}$ = $( \varnothing, v_2, v_3, v_4, v_5 )$.

$P_5$ : $\Pi_5^{Frst}$ = $( v_1, v_2, v_3, \varnothing, v_5 )$.

		Obeying Order Rule and Blend Rule 1,
		$P_2$ broadcasts $\Pi_2^{Scnd}$ on receiving $\Pi_3^{Frst}$ or $\Pi_4^{Frst}$ or $\Pi_5^{Frst}$, 
		$P_3$ broadcasts $\Pi_3^{Scnd}$ on receiving $\Pi_2^{Frst}$ or $\Pi_5^{Frst}$,
		$P_4$ broadcasts $\Pi_4^{Scnd}$ on receiving $\Pi_2^{Frst}$ or $\Pi_5^{Frst}$, and
		$P_5$ broadcasts $\Pi_5^{Scnd}$ on receiving $\Pi_2^{Frst}$ or $\Pi_3^{Frst}$ or $\Pi_4^{Frst}$.

		Obeying Completion Rule 2:
		$P_2$ completes with $V^F$ on receiving $\Pi_3^{Scnd}$, $\Pi_4^{Scnd}$, and $\Pi_5^{Scnd}$;
		$P_3$ completes with $V^F$ on receiving $\Pi_2^{Scnd}$, $\Pi_4^{Scnd}$, and $\Pi_5^{Scnd}$;
		$P_4$ completes with $V^F$ on receiving $\Pi_2^{Scnd}$, $\Pi_3^{Scnd}$, and $\Pi_5^{Scnd}$; and
		$P_5$ completes with $V^F$ on receiving $\Pi_2^{Scnd}$, $\Pi_3^{Scnd}$, and $\Pi_4^{Scnd}$.

		\textbf{Scenario 7.3}: $(N-2)$ processes received the \textit{initial value} of $P_{c}$ before they entered Proposals phase, but none of these $(N-2)$ processes has received the \textit{initial value} of $P_{s}$ before they entered Proposals phase.
		
		At start:
		 
		- $P_{r1}$ broadcast $\Pi_{r1}^{Frst}$ containing the \textit{initial value} of $P_{c}$ and a $ \varnothing $ vector element on the position of $P_{s}$; 
		
		- $P_{r2}$ broadcast $\Pi_{r2}^{Frst}$ containing the \textit{initial value} of $P_{c}$ and a $ \varnothing $ vector element on the position of $P_{s}$; 
		
		- $P_{r2}$ broadcast $\Pi_{r2}^{Frst}$ containing the \textit{initial value} of $P_{c}$ and a $ \varnothing $ vector element on the position of $P_{s}$; 
		
		- $P_{s}$ broadcast $\Pi_{s}^{Frst}$ containing a $ \varnothing $ vector element instead of the \textit{initial value} of $P_{c}$. 
		
		\textbf{Illustration}: Configuration with $P_1$ as $P_c$, $P_2$ as $P_s$, $P_3$ as $P_{r1}$, $P_4$ as $P_{r2}$, and $P_5$ as $P_{r3}$
	
		$P_1$ : $\Pi_1^{Frst}$ = $( v_1, v_2, v_3, v_4, \varnothing )$.
	
		$P_2$ : $\Pi_2^{Frst}$ = $( \varnothing, v_2, v_3, v_4, v_5 )$.
	
		$P_3$ : $\Pi_3^{Frst}$ = $( v_1, \varnothing, v_3, v_4,  v_5 )$.
	
		$P_4$ : $\Pi_4^{Frst}$ = $( v_1, \varnothing, v_3, v_4, v_5 )$.
	
		$P_5$ : $\Pi_5^{Frst}$ = $( v_1, \varnothing, v_3, v_4, v_5 )$.

		Obeying Order Rule,
		$P_2$ has $(N-2)$ equal $\Pi^{Frst}$ before having $(N-2)$ $\Pi^{Scnd}$, and obeying Completion Rule 1,
		$P_2$ completes with vector $V^\varnothing$.

		Obeying Blend Rule 1, 
		$P_2$ broadcasts $\Pi_2^{Scnd}$ on receiving $\Pi_3^{Frst}$ or $\Pi_4^{Frst}$ or $\Pi_5^{Frst}$. On receiving  $\Pi_2^{Frst}$,
		$P_3$ broadcasts $\Pi_3^{Scnd}$,
		$P_4$ broadcasts $\Pi_4^{Scnd}$, and
		$P_5$ broadcasts $\Pi_5^{Scnd}$.
		
		Obeying Completion Rule 2:
		$P_3$ completes with $V^F$ on receiving $\Pi_2^{Scnd}$, $\Pi_4^{Scnd}$, and $\Pi_5^{Scnd}$;
		$P_4$ completes with $V^F$ on receiving $\Pi_2^{Scnd}$, $\Pi_3^{Scnd}$, and $\Pi_5^{Scnd}$; and
		$P_5$ with $V^F$ on receiving $\Pi_2^{Scnd}$, $\Pi_3^{Scnd}$, and $\Pi_4^{Scnd}$.

		\textbf{Case 8}: Process $P_{c}$ crashed \textbf{after} broadcasting $\Pi_c^{Frst}$.
		
		Similarly to the cases with no crash: in both a crash \textbf{before} broadcasting $\Pi^{Scnd}$ has no effect on the outcome. In Case 1, the processes completing with $V^F$ receive $(N-1)$ $\Pi^{Scnd}$ and one less $\Pi^{Scnd}$ cannot change the outcome. In Case 2, all processes complete with $V^F$. Difference is the completion of $(N-1)$ processes. 
		
		\textbf{Case 9}:  Process $P_{c}$ crashed \textbf{while} broadcasting $\Pi_c^{Scnd}$.
		
		It is also similar to the two cases with no crash. In both a crash \textbf{while} broadcasting $\Pi^{Scnd}$ has no effect on the outcome except the completion of $(N-1)$ processes, as per Case 8.

		\textbf{Case 10}:  Process $P_{c}$ crashed \textbf{after} broadcasting $\Pi_c^{Scnd}$.

		It is also similar to the two cases with no crash, as per case 9.		

	\subsection{Comments}

		The assumed possibility of a process crash requires a possibility for agreement without input data from one of the system processes. 
		Such agreement lets avoid the infinite wait for input data from a process that might have crashed. Yet it allows the execution to continue with no input from a correct process. The main purpose of this phase is to ensure that avoiding the infinite wait does not at the same time prevent agreement with input data from all processes.
		
		This phase precludes two critically undesirable possibilities: i) for processes to complete the phase with multiple versions of vector $V^\varnothing$; or ii) only one process to complete the phase with vector $V^F$. Each process completes this phase with $V^\varnothing$ when it has received $(N-2)$ equal $\Pi^{Frst}$ messages, equal or not to its own $\Pi^{Frst}$. Alternatively, a correct process that cannot complete the phase with $V^\varnothing$, always receives $(N-2)$ $\Pi^{Scnd}$ messages to complete it with $V^F$. 

	\section{The Algorithm – Decision Phase}
		
		The system starts this phase with a precluded possibility for only one process to start it with $V^F$. Thus, either all processes start it with the same $V^\varnothing$ or $\ge 2 $ processes start it with $V^F$.
		
	\subsection{Cases with No Crash}	
		
		\textbf{Case 11}: $\ge 2 $ processes start it with $V^F$.
		
		After broadcasting a $\Delta$ message and having received $(N-2)$ $\Delta$ messages, every process prepares an ordered set of $N$ elements, which includes its own vector, the $(N-2)$ received vectors, and a $\varnothing$. 
		Every process having a $V^F$ in its set confidently terminates with $V^{De}$ = $V^F$ knowing that every non-crashed process also has a a $V^F$ in its set and will terminate with $V^{De}$ = $V^F$.
		When 2 processes started with a $V^F$ vector, every process that started with a $V^{\varnothing}$ vector will receive a message containing a $V^F$ vector and obeying the Update Rule will update its vector to $V^F$.
		
		\textbf{Illustration}: Borderline configuration where $P_1$ and $P_2$ are the processes starting with $V^F$ and each of them does not receive the other's messages within the first received (N-2) messages.
		
		The set prepared by $P_1$: $( V_1^F, \varnothing, V_3^{\varnothing}, V_4^{\varnothing}, V_5^{\varnothing} )$.
		
		The set prepared by $P_2$: $( \varnothing,  V_2^F, V_3^{\varnothing}, V_4^{\varnothing}, V_5^{\varnothing} )$.
		
		The set prepared by $P_3$: $( \varnothing,  V_2^F, V_3^{\varnothing}, V_4^{\varnothing}, V_5^{\varnothing} )$.
		
		The set prepared by $P_4$: $( \varnothing,  V_2^F, V_3^{\varnothing}, V_4^{\varnothing}, V_5^{\varnothing} )$.
		
		The set prepared by $P_5$: $( \varnothing,  V_2^F, V_3^{\varnothing}, V_4^{\varnothing}, V_5^{\varnothing} )$.
		
		Obeying the Update Rule,
		$P_3$ updates its vector from $V_3^{\varnothing}$ to $V_3^F$,
		$P_4$ updates its vector from $V_4^{\varnothing}$ to $V_4^F$, and
		$P_5$ updates its vector from $V_5^{\varnothing}$ to $V_5^F$.

		\textbf{Case 12}: $N$ processes start it with $V^{\varnothing}$.
		
		Once having received $(N-2)$ $\Delta$ messages, the set of $N$ elements assembled by each process contains $(N-1)$ vectors $V^{\varnothing}$ and a $\varnothing$. Every process having no $V^F$ in its set confidently terminates with $V^{De}$ = $V^{\varnothing}$ knowing that every non-crashed process also has no $V^F$ in its set and will terminate with $V^{De}$ = $V^{\varnothing}$.
		The non-existence of a $V^F$ vector in the own set after having received $(N-2)$ $\Delta$ messages indicates a possibility that the $\Delta$ message presented in the set with a $\varnothing$ might contain a $V^F$ vector. Such a scenario requires the system to complete Proposals Phase with two processes having completed it with $V^F$, which is precluded as a possibility with this algorithm, as demonstrated by Case 2.
	

	\subsection{Cases with A Crash}

		Let $P_{c}$ be the crashed process. The effect of the crash-fail is defined entirely by which stage the broadcasting of $\Delta_c$ message was at the point of time when the crash happened and the existence or not of processes that completed Proposals Phase with a $V^F$ vector.
		
		\textbf{Case 13}: Process $P_{c}$ crashes \textbf{before} broadcasting message $\Delta_c$.
		If $\ge 2 $ processes start it with $V^F$,
		$(N-1)$ processes broadcast a $\Delta$ message and terminate with $V^{De}$ = $V^F$, as per Case 11.
		If $N$ processes start it with $V^{\varnothing}$,
		$(N-1)$ processes broadcast a $\Delta$ message and terminate with $V^{De}$ = $V^{\varnothing}$, as per Case 12.

		\textbf{Case 14}: Process $P_{c}$ crashes \textbf{while} broadcasting message $\Delta_c$.
		It $\ge 2 $ processes start it with $V^F$,
		$(N-1)$ processes terminate with $V^{De}$ = $V^F$, as per Case 11.
		If $N$ processes start it with $V^{\varnothing}$,
		$(N-1)$ processes terminate with $V^{De}$ = $V^{\varnothing}$, as per Case 12.
			
		\textbf{Case 15}: Process $P_{c}$ crashes \textbf{after} broadcasting message $\Delta_c$.
		If $\ge 2 $ processes start it with $V^F$,
		$(N-1)$ processes terminate with $V^{De}$ = $V^F$, as per Case 11.
		If $N$ processes start it with $V^{\varnothing}$,
		$(N-1)$ processes terminate with $V^{De}$ = $V^{\varnothing}$, as per Case 12.

	\section{Binary Consensus in Asynchrony}

		This section demonstrated the possibility for any consensus to terminate in complete asynchrony with one possible crash. We show that a binary consensus can terminate (i.e., enter a decision configuration with a single decision state) even when the decision state is derived from data that does not allow a binary outcome.

		\textbf{Theorem 3}: \textit{In complete asynchrony with a possible process crash, if vector consensus protocol can terminate, then binary consensus is correct with a predetermined interpretation of binary "tie' as 0 or 1.}

		\textbf{Proof}

		The idea is to show that protocol's every admissible run is a deciding run with only one decision state. The proof follows from the next three lemmas:

		\textbf{Lemma 6}: 
		\textit{Processes of a correct binary consensus protocol produce decision states from atomically consistent datasets.}

		\textbf{Proof}

		Assume not. By definition, a consensus protocol is correct if every admissible run is a deciding run with only one decision state. A deciding run is an admissible run that has reached a decision configuration. A decision configuration is a configuration where all non-crashed processes have reached a decision state.

		Protocol correctness requires all correct processes to reach an atomically consistent binary outcome. This might happen by chance, i.e., with probability smaller than 1. In this case not every admissible run is a deciding run with only one decision state.

		Atomically consistent binary outcome with every deciding run requires: i) all correct processes to have produced atomically consistent individual datasets; and ii) every process to produce its binary outcome by passing its dataset to its own instance of the same deterministic procedure. Hence the assumption is incorrect.

		\textbf{Lemma 7}: \textit{Steps commutativity of binary consensus' final phase.}

		\textbf{Proof}

		Assume non-commutativity. Then different sequential order of steps must produce different results. However:

		- Binary consensus inevitably relies on atomically consistent datasets. [Lemma 6].

		- A binary decision derived from vector agreement data is the same regardless of steps' execution order [Theorem 1].

		As the alternative execution orders of binary consensus' final phase steps produce the same results, the assumption is wrong.

		\textbf{Lemma 8}: \textit{A binary consensus protocol can reach a deciding configuration with one decision state.}

		\textbf{Proof}

		Deciding configuration can be reached by agreeing on a vector dataset [Lemma 7]. Obtaining a binary value from the agreed upon vector dataset is a deterministic interpretation of data, including the interpretation of an outcome with binary 'tie'.

		\vspace{1\baselineskip}
		
		Impossibility of binary consensus in asynchrony and a crash was never demonstrated without reliance on the content of input data. The new paradigm isolates termination from that content.
		
		Using the new paradigm and the simultaneously ensured safety, liveness, and validity of consensus protocol with vector agreement in asynchrony, we have just demonstrated the possibility for binary consensus in asynchrony and a potential process crash.

	\section{Contrast: Crusader Agreement}

		In the context of a system with one possible crash-fail, Crusader agreement \cite{Dolev1981} implies the possibility for existence of two sets of processes, formed in the following manner as a result of the crash.
		A process successfully sends its message to a fraction of all processes immediately before crashing. All processes in this fraction form the first set. All processes outside it form the second set. 
		Crusader agreement allows the processes of the second set to enter a decision state with $ \varnothing $ in place of the missing initial value, which the processes were supposed to receive from the crash-failed process. This might create a false sense of similarity with our solution.
		
		The correctness condition of our system model does not allow a weakened agreement property, i.e., a consensus protocol is correct only if every admissible run is a deciding run with one decision state – the same for all processes. As Theorem 2 demonstrates, the presented algorithm ensures that all correct processes, in cases with or without a crash,
		decide exactly the same agreement value. With no crash, the input value of a slow process, or one with slow output links, can be excluded from the agreement value. Yet, in spite of that, this process unconditionally enters a decision state with the same agreement value as the rest of the processes.

	\section{Conclusion}

		We presented three novel perspectives on the FLP impossibility result: i) a new paradigm for consensus termination; ii) a consensus algorithm with proven termination, in complete asynchrony and tolerance to one process crash, with agreement about the initial input data; and iii) a proof that consensus with binary agreement is possible even with a tolerance to a process crash. 
	
		First, we claimed that the impossibility to produce agreement value does not itself prove impossibility to terminate consensus. In support we presented and proved the correctness of a new paradigm for consensus termination where the protocol terminates with agreement about the initial input data and then the agreed processes individually compute the consensus agreement value.
		This paradigm was a basis to claim that possibility / impossibility of consensus with binary agreement in asynchrony is not proven until there is no proof of possibility / impossibility of consensus with agreement about the initial input data.
	
		Next, we presented our algorithm for consensus with agreement about the initial input data in complete asynchrony with tolerance to one process crash and proved that the algorithm operates with simultaneously ensured safety, validity, and termination. Finally, we used the new paradigm and the algorithm to show that consensus with binary agreement can also be solved in complete asynchrony with tolerance to one crash. 
		
		The algorithm with proven termination is the main accomplishment of this work. Being a breakthrough that demonstrates the ability to neutralise the effects of communication asynchrony, it opens the door for similar results in the more important for practice area of partial synchrony \cite{DworkLynchStockm1988}.
		
	


\begin{acks}
	This work is partly sponsored by the Australian Federal Government through the Research and Development Tax Incentive Scheme. 

\end{acks}

	\newpage

	\bibliographystyle{ACM-Reference-Format}
	\bibliography{DPFLP}


\begin{thebibliography}{18}


\ifx \showCODEN    \undefined \def \showCODEN     #1{\unskip}     \fi
\ifx \showDOI      \undefined \def \showDOI       #1{#1}\fi
\ifx \showISBNx    \undefined \def \showISBNx     #1{\unskip}     \fi
\ifx \showISBNxiii \undefined \def \showISBNxiii  #1{\unskip}     \fi
\ifx \showISSN     \undefined \def \showISSN      #1{\unskip}     \fi
\ifx \showLCCN     \undefined \def \showLCCN      #1{\unskip}     \fi
\ifx \shownote     \undefined \def \shownote      #1{#1}          \fi
\ifx \showarticletitle \undefined \def \showarticletitle #1{#1}   \fi
\ifx \showURL      \undefined \def \showURL       {\relax}        \fi
\providecommand\bibfield[2]{#2}
\providecommand\bibinfo[2]{#2}
\providecommand\natexlab[1]{#1}
\providecommand\showeprint[2][]{arXiv:#2}

\bibitem[Bernstein et~al\mbox{.}(1987)]%
        {BernsteinHadzilacosGoodman}
\bibfield{author}{\bibinfo{person}{Philip Bernstein}, \bibinfo{person}{Vassos
  Hadzilacos}, {and} \bibinfo{person}{Nathan Goodman}.}
  \bibinfo{year}{1987}\natexlab{}.
\newblock \bibinfo{booktitle}{\emph{Concurrency Control and Recovery in
  Database Systems}}.
\newblock \bibinfo{publisher}{Addison-Wesley Publishing Company}.
\newblock


\bibitem[Dolev(1981)]%
        {Dolev1981}
\bibfield{author}{\bibinfo{person}{Dany Dolev}.}
  \bibinfo{year}{1981}\natexlab{}.
\newblock \showarticletitle{The Byzantine generals strike again}.
\newblock \bibinfo{journal}{\emph{Algorithm}}  \bibinfo{volume}{3}
  (\bibinfo{year}{1981}), \bibinfo{pages}{14--30}.
\newblock
Issue 1, 1982.


\bibitem[Dwork et~al\mbox{.}(1988)]%
        {DworkLynchStockm1988}
\bibfield{author}{\bibinfo{person}{Cyntia Dwork}, \bibinfo{person}{Nancy
  Lynch}, {and} \bibinfo{person}{Larry Stockmeyer}.}
  \bibinfo{year}{1988}\natexlab{}.
\newblock \showarticletitle{Consensus in the presence of partial synchrony}.
\newblock \bibinfo{journal}{\emph{J. ACM}}  \bibinfo{volume}{35}
  (\bibinfo{date}{April} \bibinfo{year}{1988}).
\newblock
Issue 2.


\bibitem[Fischer et~al\mbox{.}(1985)]%
        {FLP_result}
\bibfield{author}{\bibinfo{person}{Michael Fischer}, \bibinfo{person}{Nancy
  Lynch}, {and} \bibinfo{person}{Michael Paterson}.}
  \bibinfo{year}{1985}\natexlab{}.
\newblock \showarticletitle{Impossibility of distributed consensus with one
  faulty process}.
\newblock \bibinfo{journal}{\emph{J. ACM}}  \bibinfo{volume}{32}
  (\bibinfo{date}{April} \bibinfo{year}{1985}).
\newblock
Issue 2.


\bibitem[Garcia-Molina et~al\mbox{.}(1986)]%
        {HectorGM_EtAl1986}
\bibfield{author}{\bibinfo{person}{Hector Garcia-Molina},
  \bibinfo{person}{Frank Pittelli}, {and} \bibinfo{person}{Susan Davidson}.}
  \bibinfo{year}{1986}\natexlab{}.
\newblock \showarticletitle{Applications of Byzantine agreement in database
  systems}.
\newblock \bibinfo{journal}{\emph{ACM Transactions on Database Systems}}
  (\bibinfo{year}{1986}).
\newblock
Issue July 1986.


\bibitem[Gilbert and Lynch(2012)]%
        {GilbertLynch_2012}
\bibfield{author}{\bibinfo{person}{Seth Gilbert} {and} \bibinfo{person}{Nancy
  Lynch}.} \bibinfo{year}{2012}\natexlab{}.
\newblock \showarticletitle{Perspectives on the CAP theorem}.
\newblock \bibinfo{journal}{\emph{Computer}}  \bibinfo{volume}{45}
  (\bibinfo{year}{2012}), \bibinfo{pages}{30--36}.
\newblock


\bibitem[Gray(1978)]%
        {JimGray78}
\bibfield{author}{\bibinfo{person}{Jim Gray}.} \bibinfo{year}{1978}\natexlab{}.
\newblock \showarticletitle{Notes on Database Operating Systems, Operating
  Systems: An Advanced Course}. In \bibinfo{booktitle}{\emph{Lecture Notes in
  Computer Science}}. \bibinfo{publisher}{Springer-Verlag},
  \bibinfo{pages}{393--481}.
\newblock


\bibitem[Gray and Lamport(2006)]%
        {Gray_lamport}
\bibfield{author}{\bibinfo{person}{Jim Gray} {and} \bibinfo{person}{Leslie
  Lamport}.} \bibinfo{year}{2006}\natexlab{}.
\newblock \showarticletitle{Consensus on transaction commit}.
\newblock \bibinfo{journal}{\emph{ACM Transactions on Database Systems}}
  \bibinfo{volume}{31} (\bibinfo{date}{March} \bibinfo{year}{2006}),
  \bibinfo{pages}{133--160}.
\newblock
Issue 1.


\bibitem[Herlihy and Wing(1990)]%
        {Herlihy_Wing_1990}
\bibfield{author}{\bibinfo{person}{Maurice Herlihy} {and}
  \bibinfo{person}{Jeannette Wing}.} \bibinfo{year}{1990}\natexlab{}.
\newblock \showarticletitle{Linearizability: a correctness condition for
  concurrent objects}. In \bibinfo{booktitle}{\emph{ACM Transactions on
  Programming Languages and Systems}}, Vol.~\bibinfo{volume}{12}.
  \bibinfo{pages}{463–492}.
\newblock
Issue 3.


\bibitem[Lamport(1978)]%
        {Lamport_StateMachine}
\bibfield{author}{\bibinfo{person}{Leslie Lamport}.}
  \bibinfo{year}{1978}\natexlab{}.
\newblock \showarticletitle{The implementation of reliable distributed
  multiprocess systems}.
\newblock \bibinfo{journal}{\emph{Computer Networks}} (\bibinfo{year}{1978}),
  \bibinfo{pages}{95--114}.
\newblock
Issue 2.


\bibitem[Lamport(1979)]%
        {Lamport_1979}
\bibfield{author}{\bibinfo{person}{Leslie Lamport}.}
  \bibinfo{year}{1979}\natexlab{}.
\newblock \showarticletitle{How to make a multiprocessor computer that
  correctly executes multiprocess programs}.
\newblock \bibinfo{journal}{\emph{IEEE Trans. Comput.}} (\bibinfo{date}{Sep}
  \bibinfo{year}{1979}), \bibinfo{pages}{690 -- 691}.
\newblock
Issue 9.


\bibitem[Lampson(1996)]%
        {Lampson_1996}
\bibfield{author}{\bibinfo{person}{Butler Lampson}.}
  \bibinfo{year}{1996}\natexlab{}.
\newblock \showarticletitle{How to build a highly available systems using
  consensus}. In \bibinfo{booktitle}{\emph{Workshop on Distributed Algorithms
  (WDAG) 1996, London, UK}}. \bibinfo{publisher}{Springer-Verlag},
  \bibinfo{pages}{1--17}.
\newblock


\bibitem[Lampson and Sturgis(1976)]%
        {LampsonSturgis76}
\bibfield{author}{\bibinfo{person}{Butler Lampson} {and}
  \bibinfo{person}{Howard Sturgis}.} \bibinfo{year}{1976}\natexlab{}.
\newblock \bibinfo{title}{Crash Recovery in a Distributed Data Storage System}.
   (\bibinfo{year}{1976}).
\newblock
\newblock
\shownote{Technical Report, Xerox Palo Alto Research Center}.


\bibitem[Lindsay et~al\mbox{.}(1979)]%
        {BruceLindsay79}
\bibfield{author}{\bibinfo{person}{Bruce Lindsay}, \bibinfo{person}{Patricia
  Selinger}, \bibinfo{person}{Cesare Galtieri}, \bibinfo{person}{Jim Gray},
  \bibinfo{person}{Raymond Lorie}, \bibinfo{person}{Thomas Price},
  \bibinfo{person}{Franco Putzolu}, \bibinfo{person}{Irving Traiger}, {and}
  \bibinfo{person}{B. Wade}.} \bibinfo{year}{1979}\natexlab{}.
\newblock \bibinfo{title}{Notes on Distributed Databases}.
  (\bibinfo{year}{1979}).
\newblock
\newblock
\shownote{IBM Research Report RJ2571}.


\bibitem[Mohan et~al\mbox{.}(1985)]%
        {MohanStrongFink1985}
\bibfield{author}{\bibinfo{person}{Mohan~C. Mohan}, \bibinfo{person}{H.~Raymond
  Strong}, {and} \bibinfo{person}{Shell Finkelstein}.}
  \bibinfo{year}{1985}\natexlab{}.
\newblock \showarticletitle{Method for distributed transaction commit and
  recovery using Byzantine agreement with clusters of processors}. In
  \bibinfo{booktitle}{\emph{ACM SIGOPS Operating Systems Review 19(3)}}.
  \bibinfo{pages}{29--43}.
\newblock


\bibitem[Pease et~al\mbox{.}(1980)]%
        {PeaseShostakLamport80}
\bibfield{author}{\bibinfo{person}{Marshall Pease}, \bibinfo{person}{Robert
  Shostak}, {and} \bibinfo{person}{Leslie Lamport}.}
  \bibinfo{year}{1980}\natexlab{}.
\newblock \showarticletitle{Reaching agreement in the presence of faults}.
\newblock \bibinfo{journal}{\emph{J. ACM}}  \bibinfo{volume}{27}
  (\bibinfo{date}{April} \bibinfo{year}{1980}).
\newblock
Issue 2.


\bibitem[Santoro and Widmayer(1989)]%
        {SantoroEtAl1989}
\bibfield{author}{\bibinfo{person}{Nicola Santoro} {and} \bibinfo{person}{Peter
  Widmayer}.} \bibinfo{year}{1989}\natexlab{}.
\newblock \showarticletitle{Time is not a healer}. In
  \bibinfo{booktitle}{\emph{6th Symposium on Theoretical Aspects of Computer
  Science}}. \bibinfo{pages}{304–313}.
\newblock


\bibitem[Schneider(1990)]%
        {Schneider_1990}
\bibfield{author}{\bibinfo{person}{Fred Schneider}.}
  \bibinfo{year}{1990}\natexlab{}.
\newblock \showarticletitle{Implementing fault-tolerant services using the
  state machine approach: A tutorial}.
\newblock \bibinfo{journal}{\emph{Comput. Surveys}}  \bibinfo{volume}{22}
  (\bibinfo{year}{1990}).
\newblock
Issue 4.


\end{thebibliography}
	
	\clearpage



\end{document}